# Multiple-stage structure transformation of organic-inorganic hybrid perovskite $CH_3NH_3PbI_3$

Qiong Chen[1,*], Henan Liu[2,*], Hui-Seon Kim[3], Yucheng Liu[4], Mengjin Yang[5], Naili Yue[6], Gang Ren[6], Kai Zhu[5], Shengzhong (Frank) Liu[4], Nam-Gyu Park[3], and Yong Zhang[1,2,†]

[1]Department of Electrical and Computer Engineering

[2]Optical Science and Engineering graduate program

The University of North Carolina at Charlotte, Charlotte, NC 28223, USA

[3]School of Chemical Engineering and Department of Energy Science, Sungkyunkwan University, Suwon, 440-746, South Korea

[4]Key Laboratory of Applied Surface and Colloid Chemistry, National Ministry of Education; Institute for Advanced Energy Materials, School of Materials Science & Engineering, Shaanxi Normal University, Xi'an 710062, China

[5]Chemical and Materials Science Center, National Renewable Energy Laboratory, Golden, CO 80401, USA

[6]Molecular Foundry, Lawrence Berkeley National Laboratory, Berkeley, CA 94720, USA

* Contributed equally

† e-mail: yong.zhang@uncc.edu

**Abstract**

By performing spatially resolved Raman and photoluminescence spectroscopy with varying excitation wavelength, density, and data acquisition parameters, we have achieved a unified understanding towards the spectroscopy signatures of the organic-inorganic hybrid perovskite, transforming from the pristine state ($CH_3NH_3PbI_3$) to fully degraded state (i.e., $PbI_2$) for samples with varying crystalline domain size from mesoscopic scale (approximately 100 nm) to macroscopic size (cm), synthesized by three different techniques. We show that the hybrid perovskite exhibits multiple stages of structure transformation occurring either spontaneously or under light illumination, with exceptionally high sensitivity to the illumination conditions (e.g., power, illumination time and interruption pattern). We highlight four transformation stages (**Stage I – IV**, with **Stage I** being the pristine state) along a primary structure degradation path exhibiting distinctly different Raman spectroscopy features at each stage, and point out that previously reported Raman spectra in the literature reflect degraded structures of either **Stage III or IV**. Additional characteristic optical features of partially degraded materials under the joint action of spontaneous and photo degradation are given. This study offers reliable benchmark results for understanding the intrinsic material properties and structure transformation of this unique category of hybrid materials, and a straightforward method to monitor the structure degradation after the material is used in a device or characterized by other techniques. The findings are pertinently important to a wide range of potential applications where the hybrid material is expected to function in greatly different environment and light-matter interaction conditions.

# I. INTRODUCTION

As a solar cell absorber material, organic-inorganic hybrid structure $(CH_3NH_3)PbI_3$ ($MAPbI_3$) is perhaps the material that has showed the fastest efficiency growth among all the currently known absorber materials.[1-7] Since its first use in a sensitized solar cell in 2009 with a reported efficiency of 3.8%,[1] today the reported efficiency has exceeded 20%, an important milestone shared by only two other thin-film solar cell technologies based on CdTe and CIGS. However, this material system is facing a critical challenge, that is, structural stability.[8] The rapid progress in the device performance has led to explosive research activities on the material properties, in particular optical properties.[9] Unfortunately, some optical characterization work has been done in a rush without paying sufficient attention to the extreme structural instability, thus, yielded results of extrinsic to the material. Roughly speaking, if the photo-degradation threshold of an ionic compound like $Cu_2ZnSnSe_4$ is two-order in magnitude lower than covalent or mostly covalent semiconductors like Si and GaAs,[10, 11] $MAPbI_3$ is further lower by two orders more, as to be demonstrated in this work. Therefore, extra-caution is required in optical characterization. The primary intent of this work is to reveal the intrinsic optical properties and signatures of different stages of natural and photo-degradation, and thus provide the reliable information for further investigation of the material. Furthermore, the derived information provides guidance to different anticipated applications, from solar cell, sensor, laser, to photonic structure, where the hybrid material is expected to operate in vastly different environment and lighting conditions.

Organic-inorganic hybrid materials tend to appear in disordered phases.[12] There are only very limited examples of truly ordered hybrid materials that can be considered as crystals in a genuine sense, i.e., having a translational symmetry. Only in a few rare cases, the very high degree structural ordering can yield X-ray diffraction peaks and Raman lines as sharp as a high quality semiconductor like GaAs, as in some hybrid structures based on II-VI semiconductors, such as ZnTe(en)$_{0.5}$ with en = $C_2N_2H_8$.[13, 14] One key factor to have a highly ordered hybrid structure is understandably to involve only small organic molecules. Hybrid perovskite structure MAPbI$_3$ would seem to be a good candidate applying this simple intuition. However, because of the orientational disorder of the molecular cation MA$^+$, the structure is strictly speaking disordered (dynamic disorder).[15] Although MAPbI$_3$ is often referred to as having a tetragonal symmetry in the temperature region of approximately 162 – 327 K (cubic at higher and orthorhombic at lower temperature regions, respectively),[15] the symmetry should be understood in the same sense as in the case of a semiconductor alloy to be an averaged symmetry. As is well known that the alloying effect leads to linewidth broadening in many spectroscopy features in semiconductor alloys,[16] the disordering effect is expected to manifest in the perovskite hybrid in a similar manner.

Raman spectroscopy is an effective tool for probing the disordering effect as well as disorder-order transition in semiconductors.[17, 18] The inorganic counterpart of MAPbCl$_3$, CsPbCl$_3$, is known to be Raman inactive in the first order scattering when in the cubic phase,[19, 20] as is true for the cubic SrTiO$_3$.[21, 22] However, tetragonal and orthorhombic CsPbCl$_3$ become Raman active.[19, 20] A subtle but important difference is expected between the inorganic and hybrid perovskite. The disorder will relax the Raman selection for the cubic-phase, i.e., the first order Raman will become partially allowed, but in the meantime result in spectral broadening, which will manifest as a disorder activated band reflecting the phonon density of states, as what was observed in a similar system MAPbCl$_3$ for its room temperature cubic phase.[23] Even for its tetragonal phase, some phonon modes are in principle Raman allowed, but the disordering effect still dominated over the symmetry dictated effect, i.e., no distinct Raman mode was actually observed. Only for the low temperature orthorhombic phase, many distinct Raman peaks were observed, indicating that the structure became highly ordered.[23] Since MAPbI$_3$ is in tetragonal phase at room temperature, and likely more disorder than tetragonal MAPbCl$_3$ due to the large void for the molecule and higher temperature, the finding for MAPbCl$_3$ might hint that no distinct Raman mode should be observed for MAPbI$_3$ at room temperature. The situation in MAPbI$_3$ is further more complicated than in MAPbCl$_3$, because of lower environmental and photo stability of the former, which seems to be related to the larger anion for the former. It was found similarly for the II-VI based hybrids [13] that the photo-stability reduced with increasing anion size in common cation isostructures (e.g., CdX(ba)$_{0.5}$, X = S, S, and Te). Another subtle effect could be significant, that is, the dependence on the domain size in MAPbI$_3$ polycrystalline films. Even in an ordered inorganic semiconductor alloy, In$_{0.5}$Ga$_{0.5}$P for example, small domain size makes it much more challenging to observe ordering induced Raman features.[24] Given the fact that the hybrid perovskite materials have been synthesized using a wide range of techniques,[3-6, 25, 26] resulting in vast differences in crystalline domain size, interface and surface, it is of great interest to ask if these different structures share common vibrational signatures that are intrinsic to the hybrid structure of interest.

The reported Raman spectroscopy results for MaPbI$_3$ have in fact shown great diversity. The room temperature Raman data can be grouped into two categories: **Type 1** with relatively weak and broad bands at 52-69, 94, 108-119 cm$^{-1}$;[27-30] **Type 2** with stronger and well resolved peaks at 71, 94/95, 110/111 cm$^{-1}$.[31, 32] **Type 2** in fact closely resembles 2H-PbI$_2$ Raman modes at 74 (E$_g$), 96 (A$_{1g}$), and 110 or 106/113 cm$^{-1}$ (A$_{2u}$/2E$_u$).[33-37] Transition from **Type 1** to **2** was observed with either increasing excitation power or prolonged illumination.[29-31] The transition was explained as photo-degradation from MAPbI$_3$ to PbI$_2$,[29] but was indicated to be reversible and attributed to the photo-induced structural transformation of perovskite.[30] It is apparently unclear which set of these Raman features is intrinsic to MAPbI$_3$ or if they represent two structural variations of the perovskite framework. Note that these studies were performed using samples prepared by various techniques with one obvious variation, the polycrystalline domain size. Furthermore, the small variation in experimental conditions, such as the illumination power density and data acquisition time, could mean the photo-induced transition to be reversible or irreversible, keeping in mind that the measurements were often performed using micro-Raman systems usually involving high excitation density and the structural transformation often is an exponential activation process of the external perturbation. In our recent study using a moderately high excitation density, the hybrid perovskite Raman spectrum has been found to be domain size dependent, with that of the small domain resembling **Type 2**, whereas that of the large domain resembling the mixture of the two sets but both being well resolved.[38] Therefore, there is apparently considerable inconsistency, ambiguity and uncertainty between the reported data and regarding which represent the intrinsic properties of the material.

In this work, we attempt to address these questions: (1) are the above mentioned observations all intrinsic to the hybrid perovskite? (2) if yes, what are the likely underlying mechanisms for the variations? (3) if no, what should be the intrinsic spectroscopic features? To this end, we perform spatially resolved Raman/PL studies on a set of samples including polycrystalline films of different domain sizes and sources as well as single crystals, with varying excitation wavelength, density, and acquisition parameters. The direct comparison of the results from the same measurement system removes the possible uncertainty when comparing data from different publications, and thus is able to offer a unified understanding for the materials with distinctly different preparation methods and crystalline domain structures.

## II. EXPERIMENT

Three types samples were used in this work: (1) polycrystalline films synthesized by SKKU group that consist of polycrystalline domains of typical ~500 nm in size (referred to as **SKKU-p**), and isolated larger domains of ~5 - 10 μm (referred to as **SKKU-L**);[38, 39] (2) small domain polycrystalline films by NREL group (**NREL-p**);[40] and (3) macroscopic size (~mm - cm) single crystalline samples by SNU group (**SNU-c**).[26] Samples were received vacuum packed, and kept in a continuously pumped chamber, to prolong the sample lifetime (typically up to a few months). **Fig. 1** provides the basic structural information of the samples. Fig. 1(a) and (b) show the typical SEM images for the morphology information of the as-grown SKKU and NREL samples, respectively, and Fig. 1(c) a SEM image of a degraded SKKU sample that had been left under ambient condition for a few weeks and fully converted back into $PbI_2$. High-resolution TEM study of organic-inorganic materials is in general very challenging because of the low beam energy damage threshold. Previously reported TEM imaging of the $MAPbI_3$ perovskite hybrids was only able to show low magnification images without any lattice structure.[2, 41] Using a cryogenic temperature electron microscope with low beam energy suitable for biologic systems,[42] we were able to obtain lattice images of the hybrid material at both room and low temperature using polycrystalline samples of NREL. Lattice fringes of various lattice spacings were observed, although the zone axes could only be determined for one direction due to the limitation of the measurement system (details will be reported elsewhere). Fig. 1(d) shows one typical lattice images of a NREL sample with a lattice spacing of 6.64 Å which agrees with that of the (002) planes within the system calibration uncertainty.

The optical properties of the hybrid perovskite are very sensitive to the measurement conditions, particularly excitation wavelength λ, excitation power P or more relevant density D which in turn depending on the wavelength and numerical aperture (NA) of the microscope lens, and data collection time. Previously used power densities range from about 1 – 16 kW/cm$^2$ on the lower end [29, 30] to 86 – 110 kW/cm$^2$ on the higher end.[31, 32] In this work, we use excitation densities varying from about 170 W – 50 kW/cm$^2$, allowing more careful examination of the possible heating and photo-degradation effects. Often not specifically mentioned in the literature (except for Ref.[30]), the acquisition time is a significant parameter besides the excitation density. In this work, we have also examined the effects of total acquisition time (TAT, the actual data acquisition time excluding interruption time) as well as interruption time (IT) between continued data collection time (CT). Data were taken using either a 100x(NA = 0.9) or 50xL (NA = 0.5, long working distance) microscope lens on a Horiba LabRam HR800 confocal Raman microscope with a 1200g/mm grating, using a 532 and 441.6 nm laser with a full power of ~18 and ~13 mW, respectively, measured when exiting the microscope lens. The excitation density is estimated as D = P/A, where A is the area determined by the diffraction limit spot size 1.22λ/NA. The laser power was attenuated either using built in attenuators D1 – D4, approximately giving 1 – 4 order attenuation, or reducing the operation current of the laser. At D4, for the 532 (442) nm laser, the excitation densities were approximately 550 (580) or 170 (180) W/cm$^2$, respectively, with 100x or 50xL lens. The lowest excitation density used is significantly lower than the lowest level used in the previous reports, and is considered very low for the confocal measurement and comparable to that in a typical macroscopic measurement, although still much higher than solar radiation (0.1 W/cm$^2$ for AM1.5). In comparison to other PV materials like Si that will not show any photo-degradation under above $10^6$ W/cm$^2$ excitation, the hybrid has very low photo-stability, because prolonged illumination can lead to structural transformation even with 100 W/cm$^2$ level illumination. To minimize or examine heating and/or photo-degradation effect, a mechanical shutter was employed to block the laser beam in a Raman measurement: (1) when the CCD detector was not actively taking data; and (2) periodically during a long data collection cycle. Room temperature measurements were performed with continued low $N_2$ flow to slow down the sample degradation when exposed to ambient condition (no significant change for a few hours).

## III. RESULTS AND DISCUSSIONS

### A. Intrinsic optical properties probed by Raman and Photoluminescence

We first attempt to address the question: what to expect in the Raman spectrum of a pristine hybrid perovskite $MAPbI_3$? It turns out that all the Raman spectra reported in the literature show the spectroscopic signatures of some degree spontaneous and/or photo-induced structure degradation, based on our analyses of four representative types of samples described below.

#### 1. SKKU-p samples

**Fig. 2** compares Raman and photoluminescence (PL) spectra of a few SKKU samples. Fig. 2(a) and (b) include 5 representative Raman spectra from 5 samples using the 532 nm laser, where SKKU-p-1 to SKKU-p-4 were measured on a single spatial spot under D/TAT/CT/IT = $D4_{100x}/800s/0.5s/0.5s$ (except for SKKU-p-3 under $D4_{50xL}$), and SKKU-p-5 was measured under the same excitation level of SKKU-p-1 but only for 1s/spatial-point and averaging over 2420 total spatial points in an interval of 5 μm ($D4_{100x}/1s/1s$). The SKKU-p-5 measurement was to minimize any potential accumulative effect of the large total illumination time in the other cases. These samples represent different stages of spontaneous or natural degradation, corresponding to different lengths in the exposure time to air. The degradation is manifested in the evolution of two major spectral features: (1) in the vicinity of 100 $cm^{-1}$, from no visible feature above a smooth background in SKKU-p-1, to emergence of a small "bump" in SKKU-p-2, to two resolvable peaks in SKKU-p-3, and finally to multiple well-developed sharp peaks in SKKU-p-4 that has almost fully converted back to PbI$_2$; (2) a broad band ~550 cm$^{-1}$ with a valley at ~900 cm$^{-1}$ exhibits a systematic intensity reduction till completely vanishes from SKKU-p-1 to SKKU-p-4. This broad band was absent in the previous reports. The result of the additional sample SKKU-p-5 confirms that the spectrum of SKKU-p-1 can indeed represent that of a pristine sample. Clearly, the spectra of SKKU-p-1 and SKKU-p-5 are the two most closest to each other. The comparison between SKKU-p-1 and SKKU-p-4 indicates that the Raman cross sections near 100 cm$^{-1}$ are at least a factor of 15 difference between PbI$_2$ and MAPbI$_3$. Fig. 2(c) compares the PL spectra of the same samples measured from the same spot as that for the respective Raman spectrum (taken before the Raman measurement) using the 442 nm laser under D4$_{100x}$/1s/1s (D4$_{50xL}$/1s/1s for SKKU-p-3) in order to examine the PL emission not only near the band gap of the hybrid but also that of PbI$_2$. We notice that with increasing level of degradation, the hybrid related PL peak exhibits major peak intensity reduction, for over four orders in magnitude from SKKU-p-1 to SKKU-p-4, accompanying with blue shift in peak position, from ~769 nm in SKKU-p-1, which is near the hybrid bandgap,[43, 44] to ~743 nm in SKKU-p-4 (only visible under 532nm excitation). Additionally, in the most degraded sample SKKU-p-4, a weak PL peak at ~514 nm, which is close to the bandgap of PbI$_2$,[45] appears, whereas in the other samples, this emission is invisible. The Raman spectrum of SKKU-p-4, despite taken at a much lower excitation density level, resembles the **Type 2** spectrum for the hybrid material reported in the literature,[31, 32] but also nearly the same as those reported for PbI$_2$.[34, 35] Above results suggest that the Raman spectrum of SKKU-p-1 or SKKU-p-5 can be taken as that of a pristine MAPbI$_3$: *no visible peak near 100 cm$^{-1}$ but with a broad band at ~550 cm$^{-1}$*.

Fig. 2(d) compares Raman spectra of the most pristine sample SKKU-p-1 measured under three conditions consecutively: D4$_{100x}$/800s/0.5s/0.5s, D3$_{100x}$/400s/0.5s/0.5s, and D2$_{100x}$/100s/0.5s/0.5s. For this type of sample, the combination of D4$_{100x}$/800s/0.5s/0.5s appears to be a safe condition that does not result in visible change in the overall shape of the Raman spectrum, although maybe some intensity reduction of the 550 cm$^{-1}$ peak. Under the other two conditions, two peaks are resolved at ~96 and ~111 or ~93 and ~110 cm$^{-1}$, which are in fact similar to the **Type 1** spectra in the previous reports.[27-30] We note that in the Raman spectrum, despite the change near 100 cm$^{-1}$, the ~550 cm$^{-1}$ band remained after the two higher level measurements; also in the PL spectrum (not shown), the hybrid bandgap emission remained, and no emission near the PbI$_2$ band gap was observed. This situation represents a subtle structural change of the hybrid without significant conversion back to PbI$_2$.

In our preliminary study using the D2$_{50xL}$ excitation,[38] the isolated large domains were found to yield much stronger signal and also exhibit more Raman features than the small ones which now appears to be extrinsic in nature. Upon more careful and through examination, we suggest that the larger domains often contained incompletely reacted PbI$_2$, based on their Raman spectra and colors. Fig.2(e) compares the low excitation density (D4$_{100x}$/1000s/1.25s/3.75s) Raman spectra of two large domains, SKKUL-L-1 and SKKU-L-2, of which the former looked similar in color as the regular sample area but the later appeared brighter. The overall intensities are comparable to the small domain area (e.g., SKKU-p-1), but the brighter one exhibits PbI$_2$ like modes (similar to SKKU-p-4) superimposed on the intrinsic hybrid Raman spectrum. The difference could be explained as that the SKKU-L-2 was either slightly degraded or embedded with PbI$_2$ inclusions. Fig. 2(f) compares PL spectra of these two samples taken before the Raman measurement. SKKU-L-1 has higher peak intensity as well as longer peak wavelength than SKKU-L-2. In the PL spectrum of a typical semiconductor, the longer PL peak wavelength usually indicates the involvement of defects or impurities in the recombination. However, we have found that for the hybrid material the longer PL peak wavelength often indicates the material is less degraded, which will be further supported by the results of the SNU-c samples to be discussed later.

### 2. NREL-p samples

**Fig. 3** shows the Raman and PL spectra of a NREL-p sample. Fig. 3(a) compares two spectra taken under the same low excitation density and total acquisition time ($D4_{100x}/800s$) but with different CT/IT combinations, 1s/4s first then followed with 0.5s/0.5s. Clearly, the shorter interruption yielded a more apparent 100 cm$^{-1}$ feature. Although $D4_{100x}/800s/0.5s/0.5s$ was found to be "safe" for SKKU-p-1, one should be alerted that the "safe" condition depends on the sample condition. The difference seems to indicate that the small domains are more sensitive to the illumination than the large ones, possibly due to the lower thermal conductivity thus slower heat dissipation associated with the small polycrystalline domains, and/or the difference in surface effect. Fig. 3(b) shows the excitation density dependence: the ~100 cm$^{-1}$ feature becomes more and more apparent under D3/400s/0.5s/0.5s and D2/100s/0.5s/0.5s. These results are qualitatively consistent with those from the SKKU-p-1 sample, although there was no individually resolved peaks near 100 cm$^{-1}$ in this sample. After the D2 Raman measurement, the $D4_{100x}/800s/1s/4s$ measurement was repeated again, and the spectrum appeared to be nearly the same as the initial one, as shown in Fig. 3(c), indicating no significant permanent structure transformation in this case. PL was measured using both 442 nm and 532 nm laser at D4 with TAT = CT = 1s before and after the first and last Raman spectrum measurement, as shown in Fig. 3(d). We note that the PL peak wavelength of NREL-p sample (~764 nm) tends to be slightly shorter than that of the non-degraded SKKU samples. As shown in Fig. 3(d), although the PL peak positions show no or little change, the intensities have reduced significantly. We did not wait for a longer time to check if the intensity could recover, as partial recovery was reported in Ref.[30] after 12 hours.

Nevertheless, no PL emission near the PbI$_2$ bandgap was observed before and after the D2 Raman measurement. We note that in the NREL-p sample, even under D4$_{100x}$/800s/1s/4s, the ~550 cm$^{-1}$ band is not as prominent as in SKKU-p-1, rather comparable to that in SKKU-p-2, which seems to suggest that the sample was slightly degraded, possibly because the further reduced polycrystalline domain size in this sample made it even less stable. This observation might explain why the previous reports also did not show the ~550 cm$^{-1}$ band.[27-30]

### 3. SNU-c samples

We next examine the single crystal sample SNU-c with results given in **Fig. 4**. Fig. 4(a) contrasts two D4$_{50xL}$/1600s/0.5s/0.5s Raman spectra taken between one D2$_{50xL}$/200s/10s/0.2s measurement. While the D2 spectrum shows a visible ~100 cm$^{-1}$ peak, the two D4 spectra are nearly the same, indicating that the D2$_{50xL}$/100s/0.5s/0.5s illumination did not cause irreversible change. However, in Fig. 4(b), after increasing the D2 illumination time, with one D2$_{50xL}$/800s/20s/0.2s followed by another D2$_{50xL}$/100s/0.5s/0.5s, in the repeated D4 spectrum the 100 cm$^{-1}$ peak now becomes somewhat visible, indicating that the longer D2 illumination did induce some structural change. Also the D2 spectra are rather different between Fig. 4(a) and Fig. 4(b), i.e., longer D2 illumination has resulted in more prominent ~100 cm$^{-1}$ peak, much reduced ~550 cm$^{-1}$ band, and appearance of a broad peak near 250 cm$^{-1}$. In fact, the D2 spectra in Fig. 4(b) are similar to that in Fig. 3(b) for the NREL-p sample. Fig.4(c) shows the same measurements of Fig. 4(a) but the measured spot was flashed by D1 after the first D4 measurement. Now even at D4 major irreversible structure change is evident with the appearance of multiple peaks related to PbI$_2$. Fig. 4(d) compares the PL spectra excited by 442nm under D4$_{50xL}$/1s/1s before and after the D4 Raman of Fig. 4(b) or Fig. 4(c). After the moderately strong illumination D2$_{50xL}$/800s/20s/0.2s, the PL spectrum shows a small redshift and small intensity enhancement, indicating the material remained in the hybrid structure despite some subtle structural change; whereas after being flashed with D1, the PL spectrum reveals a factor of 88 reduction in the peak intensity and a 24 nm redshift in the peak position, indicating major structure modification. To remove the uncertainty of the extremely weak signal under the D4 excitation, another test was done under D2. The results shown in Fig. 4(e) indicates that the ~100 cm$^{-1}$ peak only appeared under D2$_{50xL}$/200s/10s/0.2s but not under D2$_{50xL}$/100s/0.5s/0.5s, with the presence of the 550 cm$^{-1}$ band in both cases. These two spectra resemble, respectively, those of SKKU-p-1 and SKKU-p-2 in Fig. 2(a). The results of this single crystal sample again confirm the conclusion about the intrinsic Raman spectrum of the material derived from the polycrystalline samples.

The results shown above offer a consistent understanding of the intrinsic Raman characteristic of the MAPbI$_3$ hybrid structure. Although the Raman modes associated with the PbI$_2$ cluster in the hybrid are expected to appear in the similar spectral range as in the bulk PbI$_2$ with the same nearest neighbor configuration (PbI$_6$), as shown theoretically,[27, 31] the Raman cross sections for these modes appear to be very small, possibly due to the symmetry selection rule,[22] as well as the inhomogeneous broadening, compared to the higher frequency modes that are likely associated with the MA molecules.[23, 27, 31] Therefore, a more comprehensive theoretical treatment including the disordering effect is required for interpreting the intrinsic Raman spectrum.

Turning to the PL spectra, our results have clearly revealed a significant and systematical variation in the peak position between the four sample types, as evident by the direct comparison of their PL spectra in Fig. 4(f). The trend seems to be that the peak wavelength red shifts with increasing domain size, which is qualitative consistent with the previously reported difference between the "meso" and "flat" sample (which differed in domain size).[28] However, since in none of these cases, the domain size was smaller enough for the quantum confinement effect to be significant, the peak wavelength shift is rather interesting. Since the small domain samples clearly exhibited strong structural disorder, i.e., no visible excitonic absorption peak even at low temperatures,[43] the PL peak position and lineshape largely reflects the density of states of the system. For the large domain or single crystalline samples, the excitonic bandgaps could be unambiguously identified at low temperatures.[43, 44] However, at room temperature it has been smeared out due to inhomogeneous broadening and could only be estimated to be around 1.63 - 1.64 eV or 760 - 756 nm,[26, 28, 43, 44] suggesting that significant disorder remains even in the large domain samples, as implied by the pseudo-tetragonal structure.[15] The room temperature PL linewidths are in the range of 80 – 100 meV with a nearly symmetric energy distribution, which is significantly different from a semiconductor like GaAs exhibiting an asymmetric PL lineshape with the higher energy side approximately following a Boltzmann distribution (shown as an inset).[46] In these hybrid samples, the PL peak positions are Stock-shifted with respect to the excitonic bandgap, as in a semiconductor alloy with strong energy fluctuation. Therefore, using the PL peak position or Tauc plot, in particular the latter, to determine the bandgap is either ambiguous or misleading.[9, 47]

### B. Photo-stability: Reversible and irreversible change, and metastable state

Photo-stability of MAPbI$_3$ depends not only on the illumination power density, total illumination time, interruption, but also on the crystalline domain size. Below, we perform more examination on the photo-stability and reversibility of the photo-induced structure changes.

#### *1. Multiple stages of photo-induced structure transformation*

The structural transformation appears to have multiple stages. **Stage I** is the as-grown or pristine structure as revealed by the Raman spectrum of SKKU-p-1 or SKKU-p-5 in Fig. 2(a) or of SNU-c in Fig. 4(e) under D2$_{50xL}$/100s/0.5s/0.5s. This stage can only been observed under very carefully controlled measurement conditions, typically very low excitation density and/or short illumination time, also depending on the sample condition (e.g., domain size). It has largely been overlooked in previous studies. Under light illumination, the material can easily be transformed into **Stage II**, which is indicated by the appearance of the "bump" near 100 cm$^{-1}$ and simultaneous reduction of the 550 cm$^{-1}$ band, as represented by the spectrum of SKKU-p-2 in Fig. 2(a), of SKKU-L-1 in Fig. 2(e), of NREL-p in Fig. 3(a), and of SNU-c under D2$_{50xL}$/200s/10s/0.2s in Fig. 4(a). Note that the material can enter this stage either due to slight natural degradation or photo-induced transformation. There is a delicate difference between the natural degradation and photo-transformation: namely for the latter, the material is in an excited state, which is less stable than the ground state. Even this stage was rarely observed in the literature, perhaps with an exception of Ref.[29], although there without the 550 cm$^{-1}$ band. In **Stage III**, distinct Raman peaks become observable near 100 cm$^{-1}$, as shown by SKKU-p-3 in Fig. 2(b) and SKKU-p-1 under higher excitation densities in Fig. 2(d) or SKKU-L-2 in Fig. 2(e). The Raman features of this stage seem to be in general agreement with those reported in the literature, such as Refs.[27-30], despite some variations in the exact peak positions, which are sensitive to the sample condition as well as the measurement condition. In **Stage IV**, the typical Raman spectrum is like that of SKKU-p-4 in Fig. 2(b), which is essentially the same as that of PbI$_2$ and also very similar to those reported hybrid spectra measured under high excitation density.[31, 32] **Stage I** appears to be a stable state at room temperature, although likely with a very small formation energy. The photo-induced transformation from **Stage I** toward **II** has been shown to be reversible, as shown in Fig. 4(a), if the illumination power and time are very carefully controlled. There might be a small energy barrier between **Stage I** and **II**. **Stage II** appears to be a metastable state, in the sense that it can endure some change towards **Stage III** yet reversible as long as not reaching **Stage III**, as to be discussed below. However, the photo-induced transformation from **Stage II** to **III** and thereafter are irreversible.

The photo-transformation is accumulative with consecutive illuminations, as demonstrated by the results shown in **Fig. 5**. Fig. 5(a) depicts 17 spectra of a SKKU-p sample, which was not been illuminated before, measured under $D2_{50xL}$/10s/0.5s/0.5s with ~20s interval between the adjacent measurements, except for #0 under $D2_{50xL}$/1.5s/0.5s/0.5s. The #0 spectrum indicates the sample was already in **Stage II** at the beginning. With increasing the repeating number, the measured point gradually evolved into **Stage III**. Accompanying the appearance of multiple resolved peaks near 100 cm$^{-1}$, the 550 cm$^{-1}$ band, as well as another band near 1340 cm$^{-1}$ that always accompanies the 550 cm$^{-1}$ band, systematically reduced. At the end of these Raman measurements, PL spectra were measured under $D4_{50xL}$/1s/1s and $D3_{50xL}$/1s/1s, as shown in Fig. 5(b). In comparison with the PL spectrum (similar to that of SKKU-p-1 in Fig. 2(c)) measured before the Raman measurements, the later PL spectra exhibit significant blue shift and major intensity reduction in the hybrid bandgap peak, and the appearance of the PbI$_2$ bandgap emission. The photo-degradation effect is qualitatively similar to the natural degradation, as shown in Fig. 2(c), but as an accelerated process. Typically, once the material entering **Stage III** under illumination, the transformation becomes irreversible. To confirm this, two additional tests were carried out: with respectively 8 or 22 consecutive measurements under $D2_{50xL}$/20s/10s/0.2s with ~20s interruption in between, then waited for 10 minutes to take the final measurement (#9 or #23, respectively). The results of the first test are shown in Fig. 5(c), indicating the material was mostly recovered (#1 vs. #9), despite some change was induced during the test, for instance, the intensity variation of the 550 cm$^{-1}$ peak between #1 and #7. However, the second test yielded irreversible change from **Stage II** to **III**, as shown in Fig. 5(d), including the spectra of #1, #6, #20, and #23, where starting from #6 individual

Raman peaks emerged in the 100 cm$^{-1}$ region. Note that there were some variations in the photo-degradation threshold between the individual spots, reflecting inhomogeneity of the polycrystalline film.

The above described photo-transformation has also been observed in the single crystalline sample SNU-c in a qualitatively similar manner. **Fig. 6** shows the illumination time and pattern dependence and spatial variation of a SNU-c sample. Fig. 6(a) and Fig. 6(b) contrasts two extreme situations due to sample inhomogeneity observed from two spatial points each with four consecutive measurements. The first and last measurement under D2$_{50xL}$/100s/0.5s/0.5s alone usually would not lead to permanent change, according to the results of Fig. 4(a) and Fig. 4(c), thus were used as probing measurements. The two longer measurements in between (total 600s illumination time) were used to induce change. In one case, Fig. 6(a), the first and the last spectrum, as well as the two middle ones, are nearly the same. The other relatively rare case, Fig. 6(b), the 100 cm$^{-1}$ region shows major enhancement and splitting starting from the third measurement, reflecting from **Stage II** to **III** transformation. Fig. 6(c) offers another test of three consecutive measurements using one single longer measurement in the middle, D2$_{50xL}$/800s/0.5s/0.5s, and the change in the last spectrum is significantly more pronounced than in Fig. 6(b). Fig. 6(d) further shows another test: after an initial D2$_{50xL}$/100s/0.5s/0.5s, followed with 8 alternating D2$_{50xL}$/100-400s/10s/0.2s and D2$_{50xL}$/100s/0.5s/0.5s measurements. Up to #7 (total illumination time of 900s from #2-#7), the change was rather small; however, another D2$_{50xL}$/800s/10s/0.2s resulted in PbI$_2$ like Raman features similar to that of D2$_{50xL}$/800s/0.5s/0.5s in Fig. 6(c), and the last probing measurement under D2$_{50xL}$/100s/0.5s/0.5s yielded a spectrum closer to that of PbI$_2$ as SKKU-p-4 in Fig. 2(b). The results here indicate that under this particular excitation density, for a single crystalline sample, approximately 800s illumination can result in irreversible structural degradation, and the effects of two different illumination patterns 800s/0.5s/0.5s and 800s/10s/0.2s are comparable.

## 2. *Photo-stability and reversibility in photoluminescence*

PL is often found more sensitive to the illumination than Raman, thus more challenging to control. Here we examine more closely the light induced changes in PL and reversibility of the effects, as well as the sample inhomogeneity using a SKKU sample. It was reported that under weak illumination (0.1 W/cm$^2$) PL initially increased slightly then decreased very significantly, but recovered partially after 12 hours.[30] Based on the above discussions, the sample there was likely already in **Stage III**. We can now investigate the PL photo-stability of the pristine (**Stage I**) sample as well as in the degraded stages under different illumination conditions.

*(a) Brief high power illumination*.   We examine the effect of a brief moderate high power (D3 or D2) illumination by measuring the low power PL before and after the flash. The response on a non-degraded SKKU sample was found to be rather non-uniform: the change in peak intensity varied roughly from +60% to -60%. If the intensity was found reduced, the same measurement was repeated again after 5 minutes. Two typical examples are given in Fig. 7(a) and (b), where the PL spectra were measured with the 532 nm laser under 14% of the $D4_{100x}$ power (14%-$D4_{100x}$/1s/1s), before and after a flash (about 1s) under 14% of the $D3_{100x}$ power. In Fig. 7(a), the change is minimal for the immediate repeat (~20s delay), whereas in Fig. 7(b) the immediate repeat exhibits a major reduction, but nearly full recovery after 5 minutes. However, not all reduced point could recover fully to the original intensity. These results indicate some delicate change in the material, although it may not necessarily be structure change, for instance, could be related to change in trapping states due to structural defects, which is worthy of further study. We note a significant difference between the current work and the previous report where the PL recovery time was found to be much longer (90% recovery after 12 hours),[30] which is likely due to the difference in sample condition (as hinted by the differences in Raman spectra) as well as the excitation density (much lower in the current study).

*(b) Prolonged low power illumination*. We examine the effects of the relatively long Raman measurement on PL or multiple repeated PL measurements. Fig. 7(c) and (d), respectively, compare the $D4_{100x}$/1s/1s PL spectra measured before and after the Raman spectra of SKKU-L-1 and SKKU-L-2 shown in Fig. 2(e) (under $D4_{100x}$/1000s/1.25s/3.75s). For SKKU-L-1, with its Raman spectrum rather close to that of a pristine sample, the intensity of the hybrid peak was found to increase by about 108%, and no emission near the $PbI_2$ bandgap was visible. However, for SKKU-L-2, with its Raman spectrum showing weak $PbI_2$ features, the intensity of the hybrid peak was found to be about 50% weaker than the other sample before the Raman measurement, and to reduce further by about 57% after the Raman measurement, with a vestige of emission near the $PbI_2$ bandgap. It is apparent that the photo-stability depends on which degradation stage the sample is. We have further examined the PL stability under repeated low power measurements and the effect on the excitation density. For instance, on one location similar to that in Fig. 7(a), we found that consecutive 8 measurements under 14% of D4 (14%-$D4_{100x}$/1s/1s) yielded 20% reduction in the peak intensity, as shown in Fig. 7(e), whereas under 10% of D4 (10%-$D4_{100x}$/1s/1s), the reduction was only 10%. However, the PL photo-stability was found to be significantly different in the SNU-c sample from that in the SKKU samples. In SNU-c, the PL intensity was found to increase with repeated measurements till reaching a saturated vale for nearly all measured locations, despite some variation in the overall intensity. Fig. 7(f) depicts the spectra of multiple consecutive measurements on one single spot under 532nm-$D4_{50xL}$/1s/1s, with intervals of 1 or 2 minutes, with the peak intensity vs. measurement number shown in the inset for multiple locations.

These results indicate that both Raman and PL spectra are very sensitive to the sample degradation. However, Raman modes can more directly reflect the structure transformation, whereas the PL spectrum is less easy to explain, in particular the intensity that has been found to exhibit either major enhancement or reduction. PL intensity in a semiconductor is known to be sensitive to extrinsic effects such as defects and surface condition.[10] Therefore, it should be used with caution for monitoring the structure transformation.

### *3. More characteristics of partially and fully degraded samples*

We provide some further discussions on additional characteristics of partially and fully degraded hybrid samples, and attempt to clarify the differences between our own latest and previous results. These results can help to identify the characteristic spectral features of the (partially) degraded hybrid samples. **Fig. 8** summarizes a few significant examples that are explained below.

*(a) Different degradation products.* We previously reported that there were substantial differences in Raman features between small and large crystalline domains in SKKU samples: the overall intensity was much lower for SKKU-p, and additional features were observed in SKKU-L,[38] as shown in Fig. 8(a). These spectra were obtained under $D2_{50xL}/200s/10s/0.2s$ with the 532 nm laser when the sample was first received (the color was dark). As now become apparent, these spectra likely reflect the degradation **Stage III**, more so for the large domain, because of the photo-degradation during the measurement. Interestingly, the highly degraded SKKU-L sample shows more features (e.g., 68, 101, ~143, ~242, ~316 cm$^{-1}$) than that of 2H or 4H PbI$_2$.[33, 36] Among these additional peaks, the 143 cm$^{-1}$ is from the TiO$_2$ buffer layer (to be explained later). As a matter of fact, the overall spectrum resembles that of ammonia (NH$_3$) intercalated PbI$_2$ in the region below 120 cm$^{-1}$.[37] If the spectrum reported in Ref.[37] is indeed of intrinsic to ammonia intercalated PbI$_2$, the SKKU-L spectrum in Fig. 8(a) could indicate that the ammonia intercalated PbI$_2$ is an intermediate state of the photo-degradation, which then points to a possible degradation path of the hybrid.

However, not all large domains exhibited the spectrum shown in Fig. 8(a), for instance, some did not show the 68 and 101 cm$^{-1}$ peaks. Those large domains that did and did not show these two peaks tended to result in two distinctly different spectra when further evolving into PbI$_2$. For comparison, Fig. 8(b) shows three representative Raman spectra measured from different locations on an appeared to be fully degraded SKKU sample (i.e., completely turned yellow) under the same condition as in Fig. 8(a), but exhibiting significantly enhanced overall intensities. After degradation, those large hybrid domains typically turned into large PbI$_2$ domains, although somewhat shrank in size. There are two distinctly different types of spectra for the fully degraded large domains. One resembles that in Fig. 8(a), but with reduced splitting between the two lowest frequency peaks (approximately from 6 to 3 cm$^{-1}$) and reduced intensity of the 316 cm$^{-1}$ peak, referred to as SKKU-L-T1. The 3 cm$^{-1}$ splitting matches that of E$_g$ mode undergoing 2H to 4H transformation,[33] which might suggest a transformation from ammonia intercalated PbI$_2$ to 4H-PbI$_2$. Shown as an inset, a comparison is made between a further degraded state from that in Fig. 8(a) and the final state. We note that the SKKU-L-T1 structure is stable in air, as measured 9 days later on the same domain. The other type, SKKU-L-T2, looks very similar to a small domain, SKKU-p-T3, in terms of both spectral features and their intensities. And they exhibit the same Raman features of 2H-PbI$_2$.

It is interesting to compare the PL spectra of the three cases in Fig. 8(b), as shown in Fig. 8(c) all taken under 442nm-D4$_{50xL}$/1s/1s. The large domain samples show stronger IR peaks than the small domain. They have all blue-shifted with respect to the non-degraded sample and are at least a factor of 40 to 1000 weaker in magnitude than a typical non-degraded sample. In all of them, the PbI$_2$ bandgap emission peaks are clearly visible, with that of the small domain sample being the strongest. The results seem to suggest that the large domains, even after achieving nearly full conversion to PbI$_2$, might still contain some molecules perhaps in the form of defects in the structure. The remaining 317 cm$^{-1}$ Raman feature in the SKKU-L-T1 spectrum in Fig. 8(b) could be an indicator.

*(b) Photo-transformation of partially degraded materials.* Photo-degradation path depends on the degradation state in which already the sample is due to the natural degradation before the illumination. We note that the D4 spectrum of NREL-p sample in Fig. 3(b) looks more like the slightly degraded SKKU-p-2 than the most pristine SKKU-p-1 in Fig. 2(a). The D2 or D3 spectrum of the NREL-p sample in Fig. 3(b) also appears to be different from those of D2 or D3 spectrum of SKKU-p-1 (starting from the pristine state) in Fig. 2(d). However, we can find some common characteristic spectral features in weakly degraded samples of all the three sources. Fig. 8(d) compares the spectrum of NREL-p at D3$_{100x}$/100s/0.5s/0.5s from Fig. 3(b), of SKKU-p-2 at D3$_{100x}$/400s/0.5s/0.5s, and of SNU-c at D2$_{50xL}$/800s/20s/0.2s from Fig. 4(b). One can see a new feature at around 250 cm$^{-1}$, which was probably buried in the broad band near 550 cm$^{-1}$ in the less degraded samples, such as Fig. 2(d) and Fig. 4(a), and remained visible in the partially degraded large domain spectrum in Fig 8(a). Note that this Raman band was also present in some previous reports,[27, 28] indicating that the results there likely represented partially degraded states. This comparison further supports that despite the considerable variations in material synthesis and therefore the crystalline structures, the hybrid materials do share common structure transformation paths in the qualitative level.

Fig. 8(e) shows an interesting PL spectrum of a severely degraded SKKU-L sample exhibiting a continuous emission spectrum from $PbI_2$ bandgap to hybrid bandgap, in contrast to the spectrum measured from the same spot that was initially only mildly degraded before being subjected to a $D2_{100x}/100s/0.5s/0.5s$ Raman measurement. The corresponding Raman spectra at $D4_{100x}/800s/0.5s/0.5s$ before and after photo-accelerated degradation are included as inset, which also reflects the transformation. The broad PL emission band expanding from NIR to green signifies a highly disordered transition phase between the hybrid perovskite and $PbI_2$.

Lastly in Fig. 8(f) we show a "peculiar" or atypical Raman spectrum from one spot of a SKKU-p sample measured under $D2_{100x}/100s/0.5s/0.5s$. It shows two Raman modes at 82 and 343 cm$^{-1}$ that did not appear in the other hybrid samples, together with a set of Raman modes of anatase $TiO_2$ at 146.9 ($E_g$), 397.7 ($B_{1g}$), (517.7) ($A_g$), and 636 cm$^{-1}$($E_g$).[48] The $TiO_2$ related peaks are no surprise, which exist in other degraded hybrid samples, such as those shown in Fig. 8(b) (not plotted), because the degraded samples were much more transparent for the 532 nm laser thus more Raman signals from the $TiO_2$ buffer layer could be detected. A subtle difference in the $TiO_2$ $E_g$ mode between this sample and those in Fig. 8(b) is noted: a blue shift from ~143.3 → ~147 cm$^{-1}$, which can be explained by the $TiO_2$ grain size change (approximately from > 30 nm to 8 nm).[49] The PL spectra from this spot, shown as an inset, under 442nm-D4(D3,D2)$_{100x}$/1s/1s are still like those of a typical hybrid sample, except for a blue shift in peak position compared to a non-degraded sample.

## IV. SUMMARY

Organic-inorganic hybrid perovskite $MAPbI_3$ has been shown to be exceptionally easy to undergo structure transformation when being characterized using common spectroscopy techniques. By applying diligent controls in the measurement conditions, we have been able to reveal the intrinsic spectroscopy signatures of the pristine samples, and characteristic spectroscopy features of degraded samples at different transformation stages, and shown that much of the previously reported Raman spectroscopy studies reflected partially degraded structures of different degrees. Importantly, despite a great variety of rather different Raman spectra have been observed either in this work or the literature, they can be attributed to different stages of structure transformation and along different paths, e.g., relatively slow natural degradation of the ground state, photo-induced accelerated degradation of the excited state, and combination of the two actions.

Intrinsic Raman spectrum of pristine MAPbI$_3$ (**Stage I** – a stable state with a small formation energy) should exhibit no visible discrete Raman modes in the spectral region of the primary Raman modes (around 100 cm$^{-1}$) of PbI$_2$, but a broad band peaked at around 550 cm$^{-1}$. The appearance of a hump in the 100 cm$^{-1}$ region, accompanying by the intensity reduction of the 550 cm$^{-1}$ band, is the first indication of structure transformation into the next stage (**Stage II** – a metastable state). The development from a hump to discrete Raman modes in the 100 cm$^{-1}$ region typically indicates the transformation into **Stage III**, and is irreversible. Ammonia (NH$_3$) intercalated PbI$_2$ could be one of the possible intermediate states. PbI$_2$ is usually the final product (**Stage IV**) of the transformation. However, different intermediate states of **Stage III** can lead to different final states, e.g., 2H-PbI$_2$ vs. 4H-PbI$_2$. A partially degraded sample may respond differently to light illumination from a pristine sample, for instance, the appearance of a Raman peak at around 250 cm$^{-1}$ that might not be observable if the initial state is **Stage I**. The distinct Raman signatures for different structure transformation stages provide a simple method to monitor the possible structure degradation after the hybrid material has been used in a device or characterized by other techniques.

On the PL spectrum of MAPbI$_3$, the structure transformation typically yields the intensity reduction and peak blue shift of the emission near the hybrid bandgap. A severely degraded sample tends to yield emission near the bandgap of the PbI$_2$. The polycrystalline samples tend to show inhomogeneous response to light illumination, for instance, exhibiting either enhanced or reduced PL intensity, PL quenching being either reversible or irreversible. The single crystalline sample tends to show increasing PL intensity till saturation under repeated measurements. The PL intensity change may or may not reflect the structural change. Relatively speaking, PL is a less reliable technique for monitoring the structure transformation compared to Raman that is more directly correlated with the crystalline structure.

## ACKNOWLEDGEMENT

The work at UNC-Charlotte was partially supported with funds from Y.Z.'s Bissell Distinguished Professorship; at SKKU by the National Research Foundation of Korea (NRF) grants funded by the Ministry of Science, ICT & Future Planning (MSIP) of Korea under Contract Nos. NRF-2010-0014992, NRF-2012M1A2A2671721, NRF-2012M3A7B4049986 (Nano Material Technology Development Program), and NRF-2012M3A6A7054861 (Global Frontier R&D Program on Center for Multiscale Energy System); at NREL by the hybrid perovskite solar cell program of the National Center for Photovoltaics funded by the U.S. Department of Energy, Office of Energy Efficiency and Renewable Energy, Solar Energy Technologies Office, for the work performed at the National Renewable Energy Laboratory (Contract no. DE-AC36-08-GO28308); at SNU by the National University Research Fund (GK261001009), the Changjiang

Scholar and Innovative Research Team (IRT_14R33), the Overseas Talent Recruitment Project (B14041) and the Chinese National 1000-talent-plan program (1110010341); at the Molecular Foundry by the Office of Science, Office of Basic Energy Sciences, U.S. Department of Energy, under Contract No. DE-AC02-05CH11231. Y.Z. thanks P. A. Beckmann for discussions about $PbI_2$ polytypism, J. E. Spanier for discussions about Raman selection rules in perovskite structures.

# References

[1] A. Kojima, K. Teshima, Y. Shirai, and T. Miyasaka, Organometal Halide Perovskites as Visible-Light Sensitizers for Photovoltaic Cells. *Journal of the American Chemical Society* **131**, 6050 (2009).
[2] J.-H. Im, C.-R. Lee, J.-W. Lee, S.-W. Park, and N.-G. Park, 6.5% efficient perovskite quantum-dot-sensitized solar cell. *Nanoscale* **3**, 4088 (2011).
[3] J. Burschka, N. Pellet, S.-J. Moon, R. Humphry-Baker, P. Gao, M. K. Nazeeruddin, and M. Gratzel, Sequential deposition as a route to high-performance perovskite-sensitized solar cells. *Nature* **499**, 316 (2013).
[4] M. M. Lee, J. Teuscher, T. Miyasaka, T. N. Murakami, and H. J. Snaith, Efficient Hybrid Solar Cells Based on Meso-Superstructured Organometal Halide Perovskites. *Science* **338**, 643 (2012).
[5] N. J. Jeon, J. H. Noh, Y. C. Kim, W. S. Yang, S. Ryu, and S. I. Seok, Solvent engineering for high-performance inorganic–organic hybrid perovskite solar cells. *Nat Mater* **13**, 897 (2014).
[6] W. Nie, H. Tsai, R. Asadpour, J.-C. Blancon, A. J. Neukirch, G. Gupta, J. J. Crochet, M. Chhowalla, S. Tretiak, M. A. Alam, H.-L. Wang, and A. D. Mohite, High-efficiency solution-processed perovskite solar cells with millimeter-scale grains. *Science* **347**, 522 (2015).
[7] Z. Xiao, C. Bi, Y. Shao, Q. Dong, Q. Wang, Y. Yuan, C. Wang, Y. Gao, and J. Huang, Efficient, high yield perovskite photovoltaic devices grown by interdiffusion of solution-processed precursor stacking layers. *Energy & Environmental Science* **7**, 2619 (2014).
[8] M. Gratzel, The light and shade of perovskite solar cells. *Nat Mater* **13**, 838 (2014).
[9] M. A. Green, Y. Jiang, A. M. Soufiani, and A. Ho-Baillie, Optical Properties of Photovoltaic Organic–Inorganic Lead Halide Perovskites. *The Journal of Physical Chemistry Letters* **6**, 4774 (2015).
[10] T. H. Gfroerer, Y. Zhang, and M. W. Wanlass, An extended defect as a sensor for free carrier diffusion in a semiconductor. *Applied Physics Letters* **102**, 012114 (2013).
[11] Q. Chen, and Y. Zhang, The reversal of the laser-beam-induced-current contrast with varying illumination density in a $Cu_2ZnSnSe_4$ thin-film solar cell. *Applied Physics Letters* **103**, 242104 (2013).
[12] H. S. Nalwa, *Handbook of organic-inorganic hybrid materials and nanocomposites* (American Scientific Publishers, 2003).
[13] X. Y. Huang, J. Li, Y. Zhang, and A. Mascarenhas, From 1D chain to 3D network: Tuning hybrid II-VI nanostructures and their optical properties. *Journal of the American Chemical Society* **125**, 7049 (2003).
[14] Y. Zhang, Z. Islam, Y. Ren, P. A. Parilla, S. P. Ahrenkiel, P. L. Lee, A. Mascarenhas, M. J. McNevin, I. Naumov, H. X. Fu, X. Y. Huang, and J. Li, Zero Thermal Expansion in a Nanostructured Inorganic-Organic Hybrid Crystal. *Physical Review Letters* **99**, 215901 (2007).
[15] A. Poglitsch, and D. Weber, Dynamic disorder in methylammoniumtrihalogenoplumbates (II) observed by millimeter‐wave spectroscopy. *The Journal of Chemical Physics* **87**, 6373 (1987).
[16] Y. Zhang, and L.-W. Wang, Global electronic structure of semiconductor alloys through direct large-scale computations for III-V alloys $Ga_xIn_{1-x}P$. *Physical Review B* **83**, 165208 (2011).
[17] B. Jusserand, and M. Cardona, in *Light Scattering in Solids*, edited by M. Cardona, and G. Güntherodt (Springer, Berlin, 1989), p. 49.
[18] A. Mascarenhas, H. Cheong, and F. Alsina, in *Spontaneous ordering in semiconductor alloys*, edited by A. Mascarenhas (Kluwer Academy, New York, 2002), p. 391.
[19] S. Hirotsu, Far-infrared reflectivity spectra of CsPbCl3. *Physics Letters A* **41**, 55 (1972).
[20] D. M. Calistru, L. Mihut, S. Lefrant, and I. Baltog, Identification of the symmetry of phonon modes in $CsPbCl_3$ in phase IV by Raman and resonance-Raman scattering. *Journal of Applied Physics* **82**, 5391 (1997).
[21] W. G. Nilsen, and J. G. Skinner, Raman Spectrum of Strontium Titanate. *The Journal of Chemical Physics* **48**, 2240 (1968).

[22]    M. A. Islam, J. M. Rondinelli, and J. E. Spanier, Normal mode determination of perovskite crystal structures with octahedral rotations: theory and applications. *Journal of Physics: Condensed Matter* **25**, 175902 (2013).
[23]    A. Maalej, Y. Abid, A. Kallel, A. Daoud, A. Lautié, and F. Romain, Phase transitions and crystal dynamics in the cubic perovskite $CH_3NH_3PbCl_3$. *Solid State Communications* **103**, 279 (1997).
[24]    H. M. Cheong, A. Mascarenhas, P. Ernst, and C. Geng, Effects of spontaneous ordering on Raman spectra of GaInP. *Physical Review B* **56**, 1882 (1997).
[25]    C. C. Stoumpos, C. D. Malliakas, and M. G. Kanatzidis, Semiconducting Tin and Lead Iodide Perovskites with Organic Cations: Phase Transitions, High Mobilities, and Near-Infrared Photoluminescent Properties. *Inorganic Chemistry* **52**, 9019 (2013).
[26]    Y. Liu, Z. Yang, D. Cui, X. Ren, J. Sun, X. Liu, J. Zhang, Q. Wei, H. Fan, F. Yu, X. Zhang, C. Zhao, and S. Liu, Two-Inch-Sized Perovskite $CH_3NH_3PbX_3$ (X = Cl, Br, I) Crystals: Growth and Characterization. *Advanced Materials* **27**, 5176 (2015).
[27]    C. Quarti, G. Grancini, E. Mosconi, P. Bruno, J. M. Ball, M. M. Lee, H. J. Snaith, A. Petrozza, and F. D. Angelis, The Raman Spectrum of the $CH_3NH_3PbI_3$ Hybrid Perovskite: Interplay of Theory and Experiment. *The Journal of Physical Chemistry Letters* **5**, 279 (2013).
[28]    G. Grancini, S. Marras, M. Prato, C. Giannini, C. Quarti, F. De Angelis, M. De Bastiani, G. E. Eperon, H. J. Snaith, L. Manna, and A. Petrozza, The Impact of the Crystallization Processes on the Structural and Optical Properties of Hybrid Perovskite Films for Photovoltaics. *The Journal of Physical Chemistry Letters* **5**, 3836 (2014).
[29]    M. Ledinský, P. Löper, B. Niesen, J. Holovský, S.-J. Moon, J.-H. Yum, S. De Wolf, A. Fejfar, and C. Ballif, Raman Spectroscopy of Organic–Inorganic Halide Perovskites. *The Journal of Physical Chemistry Letters* **6**, 401 (2015).
[30]    R. Gottesman, L. Gouda, B. S. Kalanoor, E. Haltzi, S. Tirosh, E. Rosh-Hodesh, Y. Tischler, A. Zaban, C. Quarti, E. Mosconi, and F. De Angelis, Photoinduced Reversible Structural Transformations in Free-Standing $CH_3NH_3PbI_3$ Perovskite Films. *The Journal of Physical Chemistry Letters* **6**, 2332 (2015).
[31]    B.-w. Park, S. M. Jain, X. Zhang, A. Hagfeldt, G. Boschloo, and T. Edvinsson, Resonance Raman and Excitation Energy Dependent Charge Transfer Mechanism in Halide-Substituted Hybrid Perovskite Solar Cells. *ACS Nano* **9**, 2088 (2015).
[32]    S. T. Ha, X. Liu, Q. Zhang, D. Giovanni, T. C. Sum, and Q. Xiong, Synthesis of Organic–Inorganic Lead Halide Perovskite Nanoplatelets: Towards High-Performance Perovskite Solar Cells and Optoelectronic Devices. *Advanced Optical Materials* **2**, 838 (2014).
[33]    R. Zallen, and M. L. Slade, Inter-polytype conversion and layer-layer coupling in $PbI_2$. *Solid State Communications* **17**, 1561 (1975).
[34]    V. Capozzi, A. Fontana, M. P. Fontana, G. Mariotto, M. Montagna, and G. Viliani, Raman scattering in $PbI_2$. *Nuov Cim B* **39**, 556 (1977).
[35]    W. M. Sears, M. L. Klein, and J. A. Morrison, Polytypism and the vibrational properties of $PbI_2$. *Physical Review B* **19**, 2305 (1979).
[36]    A. Grisel, and P. Schmid, Polytypism and Lattice Vibrations of $PbI_2$. *physica status solidi (b)* **73**, 587 (1976).
[37]    N. Preda, L. Mihut, M. Baibarac, I. Baltog, and S. Lefrant, A distinctive signature in the Raman and photoluminescence spectra of intercalated $PbI_2$. *Journal of Physics: Condensed Matter* **18**, 8899 (2006).
[38]    H. Liu, Q. Chen, H.-S. Kim, N.-G. Park, and Y. Zhang, in IEEE PVSCNew Orlean, 2015).
[39]    J.-H. Im, H.-S. Kim, and N.-G. Park, Morphology-photovoltaic property correlation in perovskite solar cells: One-step versus two-step deposition of $CH_3NH_3PbI_3$. *APL Materials* **2** (2014).
[40]    Y. Zhao, A. M. Nardes, and K. Zhu, Solid-State Mesostructured Perovskite $CH_3NH_3PbI_3$ Solar Cells: Charge Transport, Recombination, and Diffusion Length. *The Journal of Physical Chemistry Letters* **5**, 490 (2014).

[41]   A. Mei, X. Li, L. Liu, Z. Ku, T. Liu, Y. Rong, M. Xu, M. Hu, J. Chen, Y. Yang, M. Grätzel, and H. Han, A hole-conductor–free, fully printable mesoscopic perovskite solar cell with high stability. *Science* **345**, 295 (2014).

[42]   G. Ren, G. Rudenko, S. J. Ludtke, J. Deisenhofer, W. Chiu, and H. J. Pownall, Model of human low-density lipoprotein and bound receptor based on CryoEM. *Proceedings of the National Academy of Sciences* **107**, 1059 (2010).

[43]   V. D'Innocenzo, G. Grancini, M. J. P. Alcocer, A. R. S. Kandada, S. D. Stranks, M. M. Lee, G. Lanzani, H. J. Snaith, and A. Petrozza, Excitons versus free charges in organo-lead tri-halide perovskites. *Nat Commun* **5**, 3586 (2014).

[44]   A. M. Soufiani, F. Huang, P. Reece, R. Sheng, A. Ho-Baillie, and M. A. Green, Polaronic exciton binding energy in iodide and bromide organic-inorganic lead halide perovskites. *Applied Physics Letters* **107**, 231902 (2015).

[45]   I. Baltog, M. Baibarac, and S. Lefrant, Quantum well effect in bulk $PbI_2$ crystals revealed by the anisotropy of photoluminescence and Raman spectra. *Journal of Physics: Condensed Matter* **21**, 025507 (2009).

[46]   H. B. Bebb, and E. W. Williams, in *Transport and Optical Phenomena*, edited by R. K. Willardson, and A. C. Beer (Academic Press, New York, 1972), p. 239.

[47]   Z. Yong, B. Fluegel, M. C. Hanna, J. F. Geisz, L. W. Wang, and A. Mascarenhas, Effects of heavy nitrogen doping in III-V semiconductors - How well does the conventional wisdom hold for the dilute nitrogen "III-V-N alloys"? *Physica Status Solidi C*, 396 (2003).

[48]   T. Ohsaka, F. Izumi, and Y. Fujiki, Raman spectrum of anatase, $TiO_2$. *Journal of Raman Spectroscopy* **7**, 321 (1978).

[49]   W. F. Zhang, Y. L. He, M. S. Zhang, Z. Yin, and Q. Chen, Raman scattering study on anatase $TiO_2$ nanocrystals. *Journal of Physics D: Applied Physics* **33**, 912 (2000).

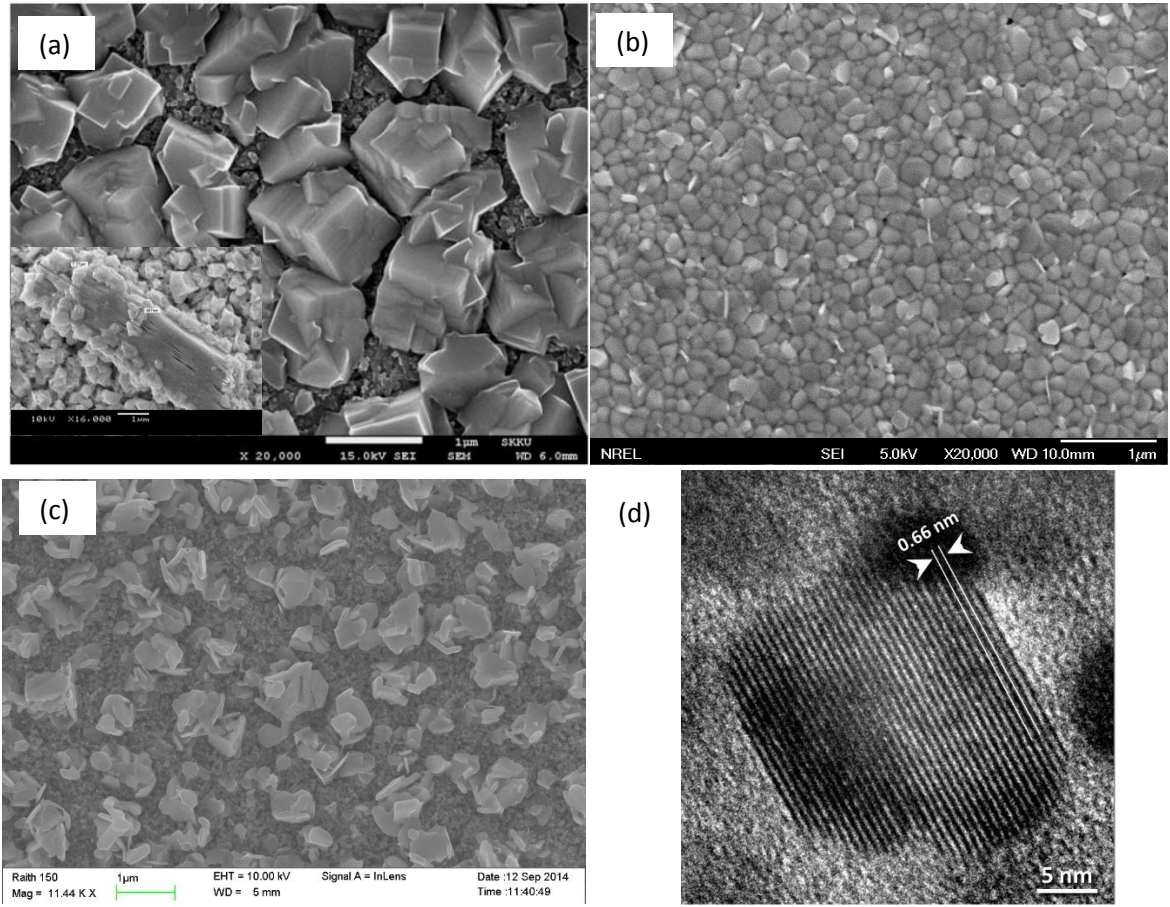

**FIG. 1.** SEM and TEM images of MAPbI$_3$ hybrid perovskite samples. (a) – (c) SEM images of a SKKU-p sample (with a large domain shown as an inset), NREL-p sample, and degraded SKKU-p sample. (d) A TEM image of a NREL-p sample.

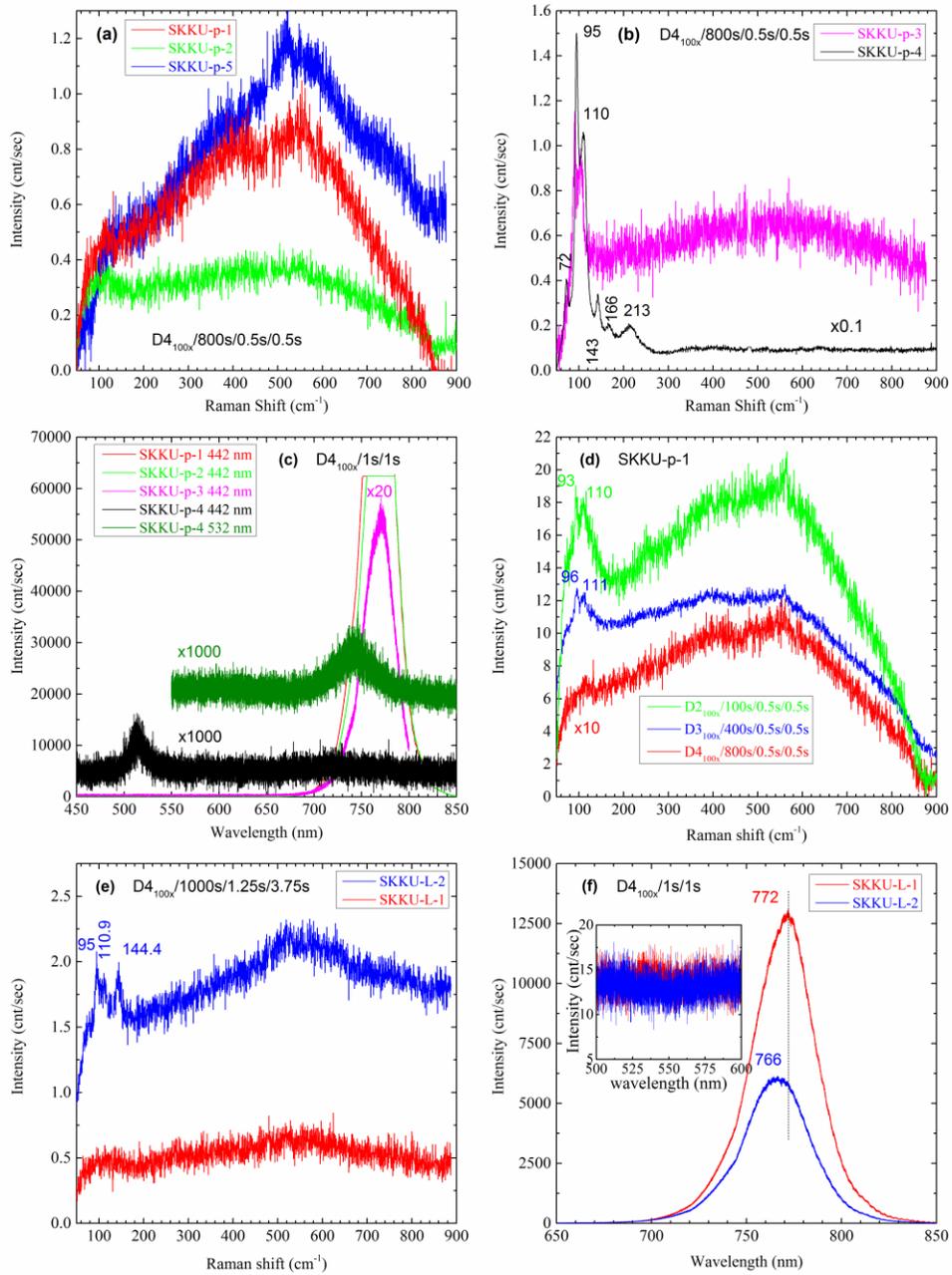

**FIG. 2.** Raman and PL spectra of SKKU samples at different stages of structure transformation. (a) & (b) Representative Raman spectra at different stages of structure transformation from $MaPbI_3$ to $PbI_2$, measured under a very low excitation density D4 (SKKU-p-3 was measured with the 50xL lens, but multiplied by the laser density ratio between the 100x and 50xL lens). (c) PL spectra measured at D4 from the same location as in (a) & (b) before the Raman measurement. (d) Raman excitation density dependence of the most pristine sample SKKU-p-1. (e) Low excitation density (D4) Raman spectra for two large domain SKKU-L samples. (f) The corresponding PL spectra of (e), measured under D4 before the Raman measurement. The inset shows the spectral region near the $PbI_2$ bandgap.

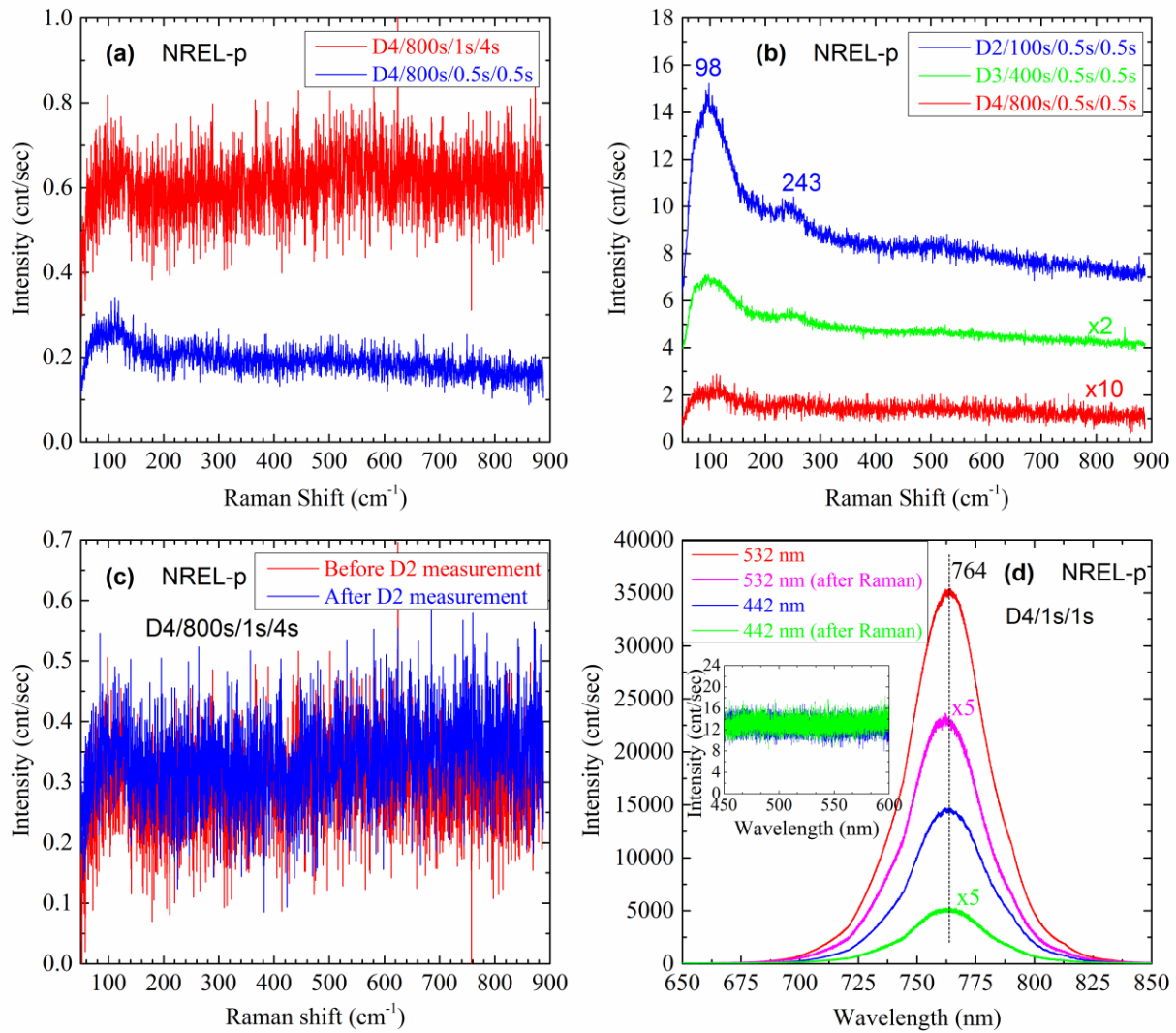

**FIG. 3.** Raman and PL spectra of NREL-p samples measured at different conditions. (a) D4 Raman spectra measured with different interruptions. (b) Raman spectra measured under different excitation density D4-D2. (c) D4/800s/1s/4s measurements before and after the measurements shown in (b). (d) PL spectra measured before and after all the Raman measurements, using both 442 and 532 nm laser.

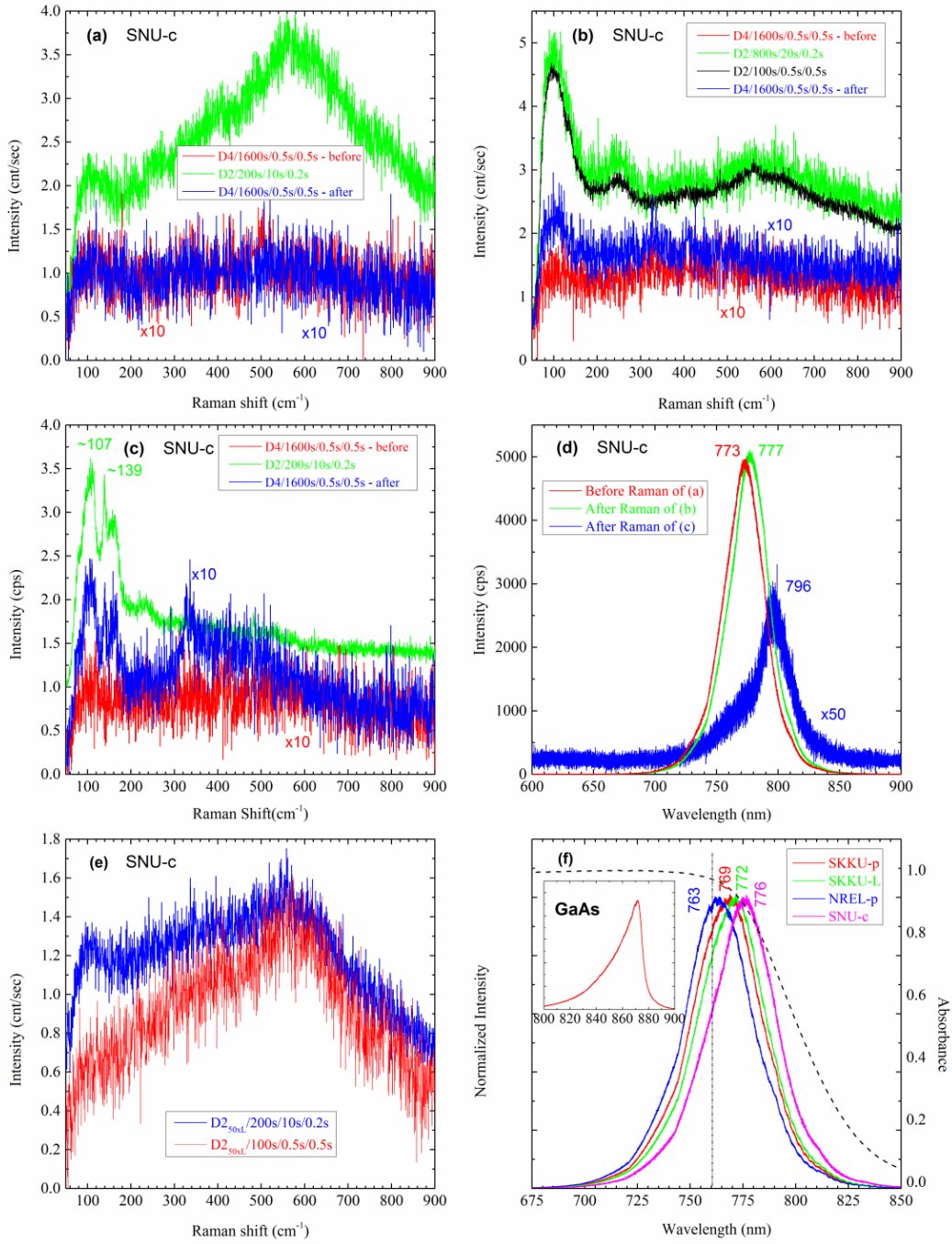

**FIG. 4.** Raman and PL spectra of SNU-c samples measured at different conditions. (a) - (c) Low excitation density (D4) Raman spectra measured before and after different higher excitation density measurements. (d) D4 PL spectra measured before the Raman measurement shown in (a) and after the Raman measurements shown in (b) and (c). (e) Raman spectra measured at a moderately high excitation density D2 under different data collection conditions. (f) Comparison of the D4 PL spectra for four types of samples, SKKU-p, SKKU-L, NREL-p, and SNU-c, and an absorption spectrum of a SNU-c sample. The inset shows a room temperature GaAs PL spectrum.

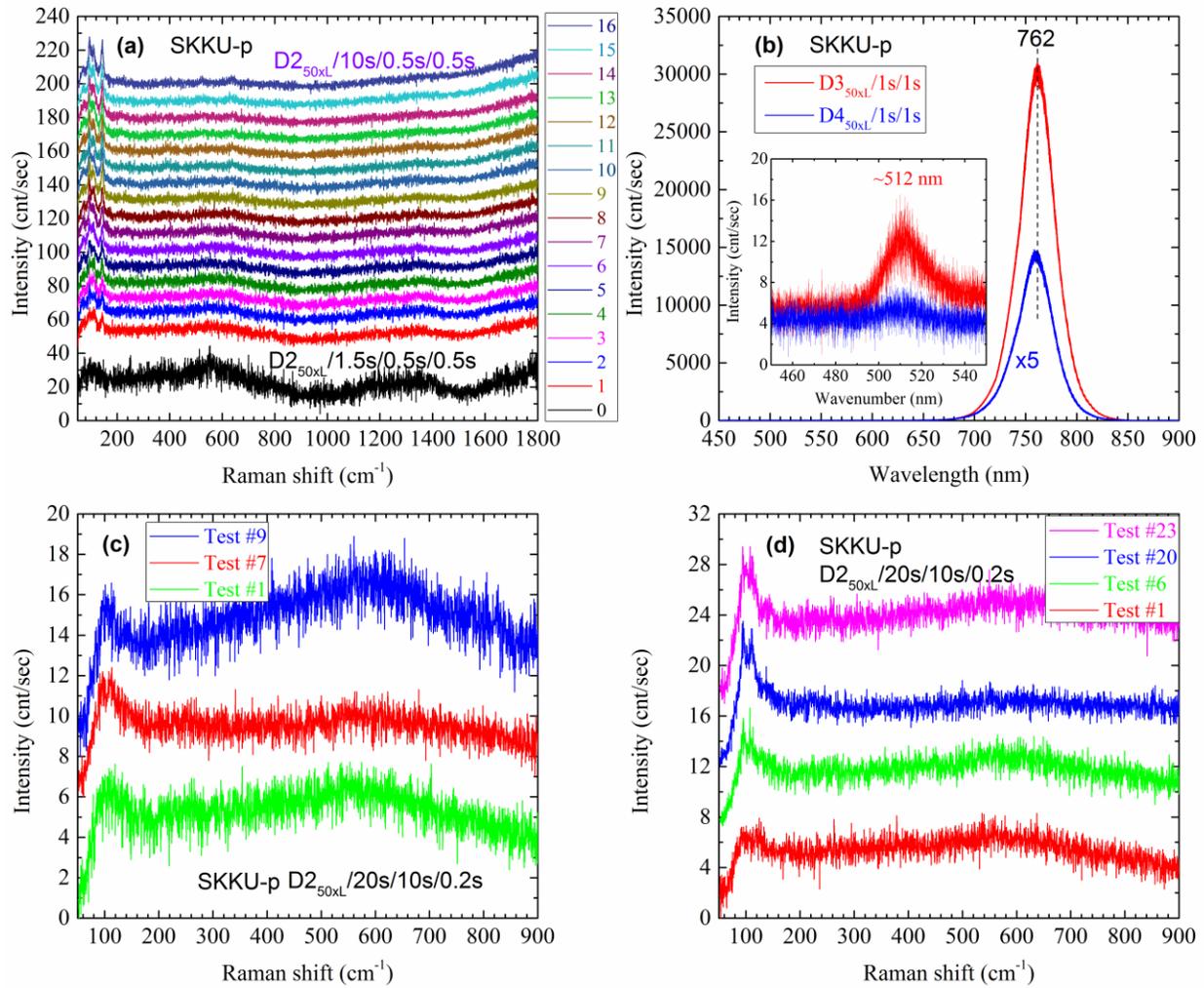

**FIG. 5.** Effects of multiple consecutive Raman measurements in SKKU-p samples under D2 excitation. (a) 16 consecutive measurements with ~20s in between. (b) PL spectra measured after the Raman measurements shown in (a). (c) 8 consecutive measurements with ~20s in between, then waited for 10 minutes before the last one (#9). (d) 22 consecutive measurements with ~20s in between, then waited for 10 minutes before the last one (#23).

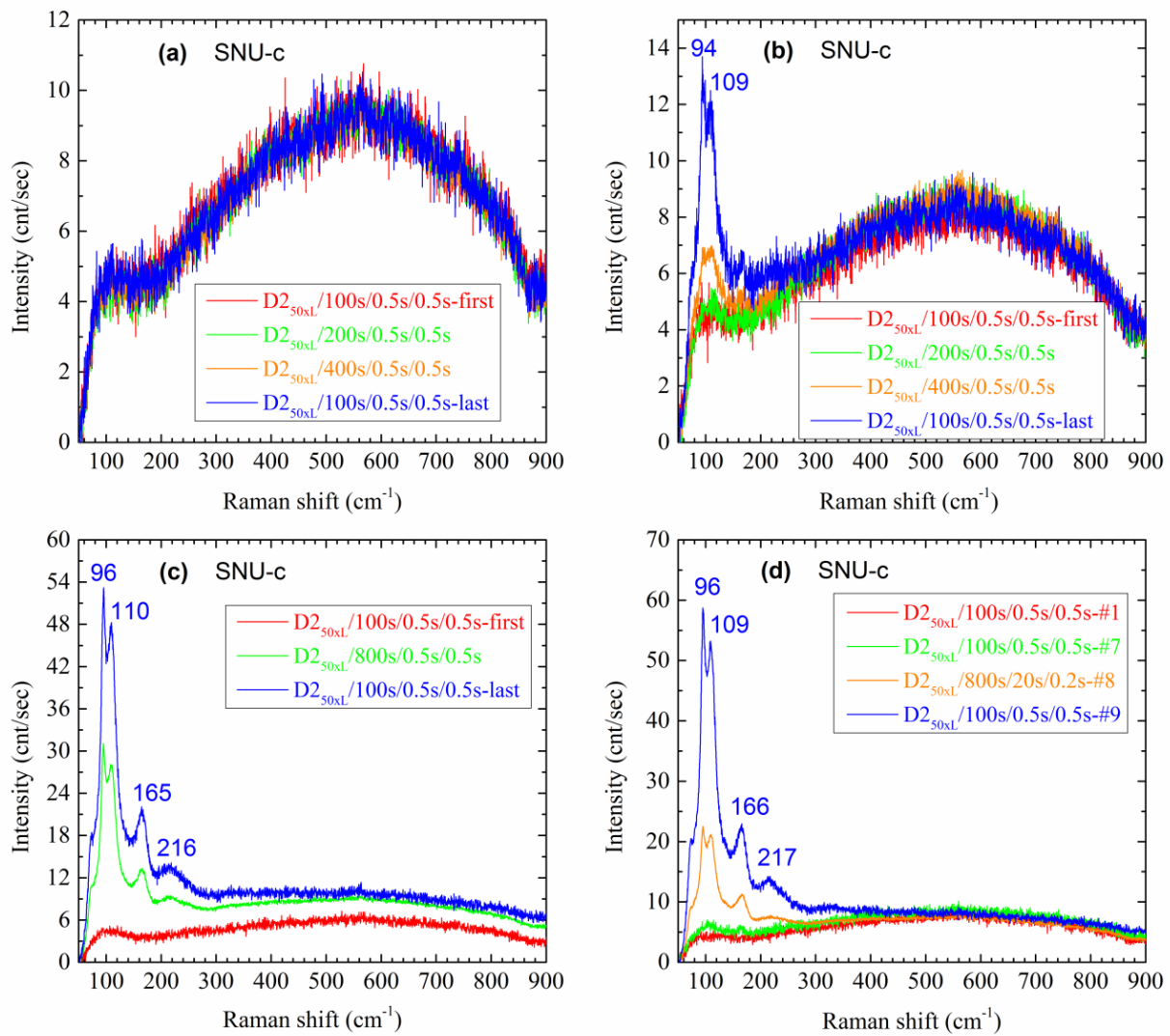

**FIG. 6.** Effects of multiple consecutive Raman measurements in SNU-c samples under D2 excitation. (a) and (b) show spatial inhomogeneity. (c) and (d) show the effect of long illumination time.

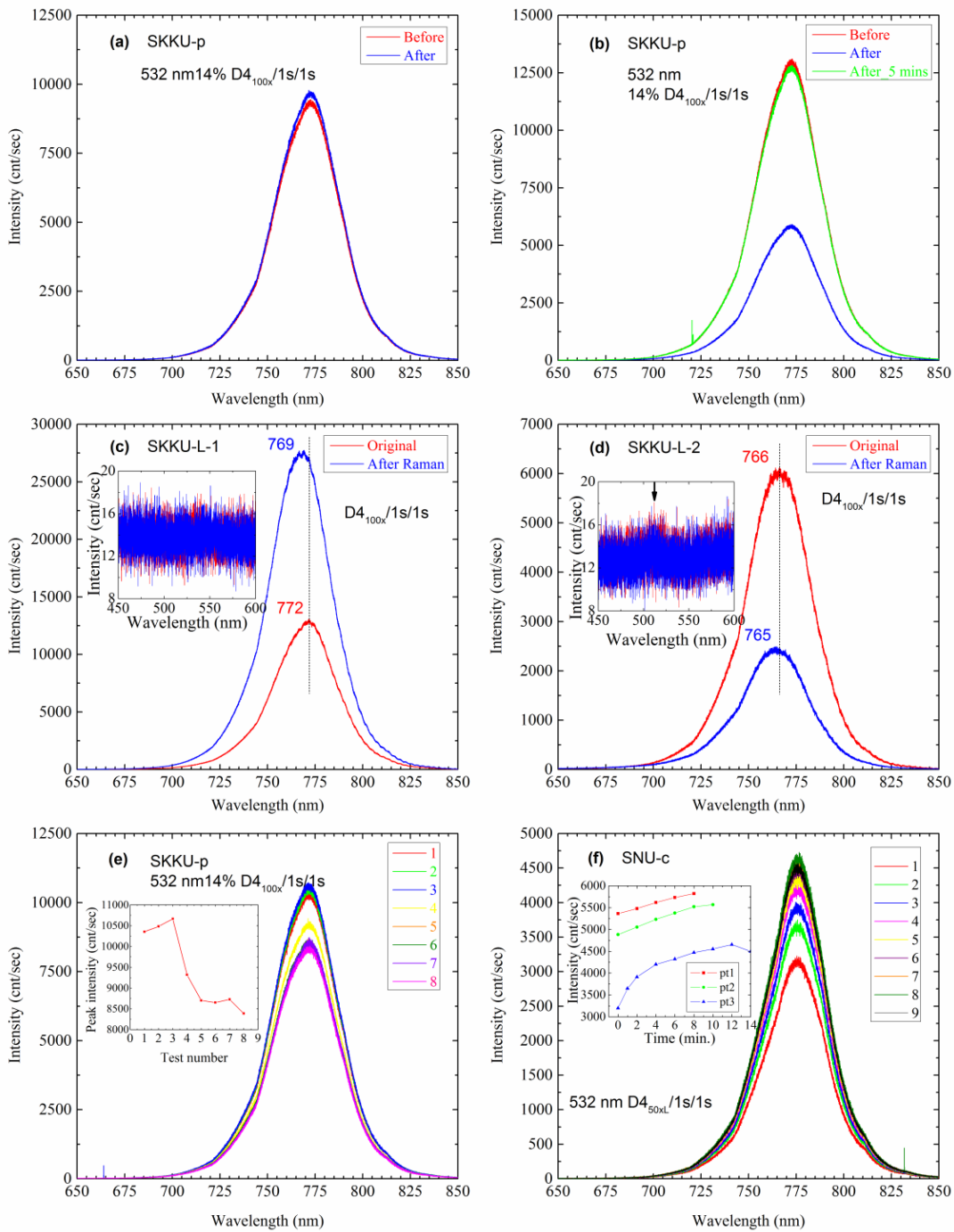

**FIG. 7.** Photo-stability of the PL spectrum. (a) and (b) for two SKKU-p samples, low excitation density PL spectra before and after a higher power flash (~1s). (c) and (d) for two SKKU-L samples, the same as those yielded the spectra in Fig. 2(e) and (f), PL spectra measured before and after the Raman measurements shown in Fig. 2(e). (e) Consecutively measured PL spectra of a SKKU-p sample under very low excitation density. (f) Consecutively measured PL spectra of a SNU-c sample under very low excitation density.

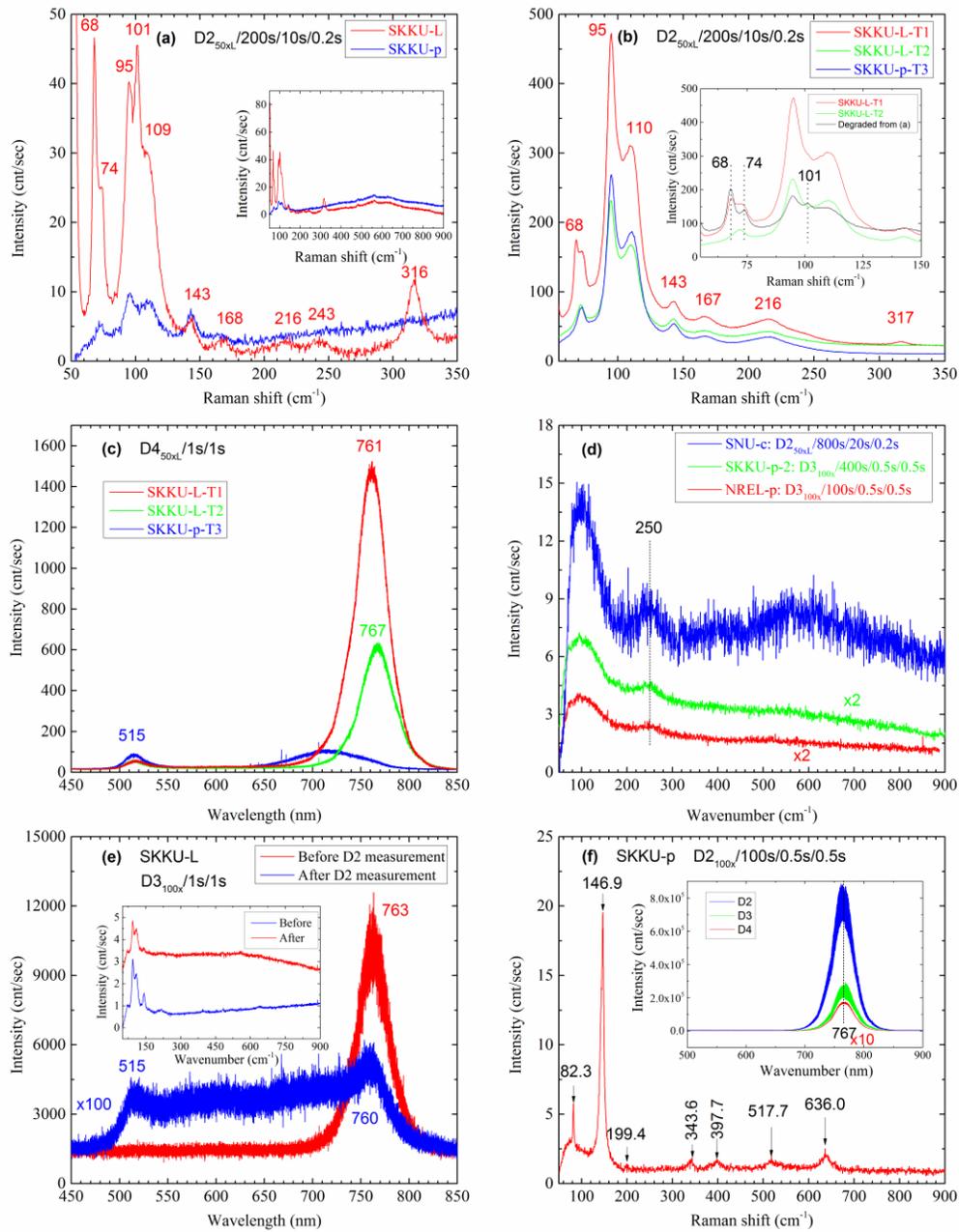

**FIG. 8.** Raman and PL spectra of partially or fully degraded samples. (a) Comparison of partially photo-degraded SKKU sample between large and small domain (the inset shows the spectra in a larger spectral range). (b) Nearly fully degraded SKKU samples (including two distinctively different large domains), and the inset shows the spectra in a smaller spectral range. (c) PL spectra of the samples shown in (b). (d) The effect of photo-induced structure transformation in an already partially degraded sample due to nature course, showing a common Raman peak at around 250 cm$^{-1}$. (e) PL spectra of a partially degraded SKKU-L sample measured before and after high excitation density Raman measurement (the inset shows the corresponding Raman spectra). (f) Raman and PL spectra of an atypical location in a SKKU-p sample.